\documentclass[11pt]{article}
\usepackage[utf8]{inputenc}
\usepackage{fullpage}
\usepackage{amsfonts}
\usepackage{amsmath}
\usepackage{amssymb}
\usepackage{bm}
\usepackage{cite}
\usepackage{graphicx}
\usepackage{subcaption}
\usepackage[dvipsnames]{xcolor}
\usepackage[hidelinks]{hyperref}

\title{Stochastic Heat Engine Using Multiple Interacting Active Particles}
\author{Aradhana Kumari$^1$\footnote{aradhanakumari2546@gmail.com}, Md Samsuzzaman$^2$\footnote{samsuzz@gmail.com}, Arnab Saha$^3$\footnote{sahaarn@gmail.com}, and Sourabh Lahiri$^1$\footnote{sourabhlahiri@bitmesra.ac.in}}
\date{}

\begin{document}

\maketitle

\begin{center}
    $^1$Department of Physics, Birla Institute of Technology Mesra, Jharkhand 835215, India \\
    $^2$Department of Physics, Savitribai Phule Pune University, Pune 411007, India\\
    $^3$Department of Physics, University Of Calcutta, 92 Acharya Prafulla Chandra Road, Kolkata-700009, India\\
\end{center}

\begin{abstract}
    The area of stochastic heat engines using active particles has attracted a lot of attention recently. They have been shown to exhibit advantages over engines using passive particles. In this work, we use multiple self-propelling particles undergoing Vicsek-like aligning interaction as our working system. The particles are confined in a two-dimensional circular trap. The interplay between the confinement and the activity of the particles induces clustering. These clusters change their locations relative to the walls of the trap, when the wall steepness is varied with time. In this work we demonstrate that changing the steepness of the wall and the activity of the particles time-periodically can cause the system to act as an engine. In this setup, we study the variations in extracted work with the activity, rotational diffusion, and the Vicsek radius of individual particles. We also comment on the complications involved in the definition of the engine efficiency in accordance with the usual prescription of stochastic thermodynamics.
\end{abstract}

\section{Introduction}

The study of heat engines has been the cornerstone of our understanding of thermodynamics. In fact, almost the entire modern thermodynamics that we are familiar with originated from the early studies of Carnot on heat engines of high efficiencies \cite{Carnot1824,Carnot1897,Callen}. Since then, these devices have remained one of the most useful tools in everyday life, obvious examples being automobiles, pumps, firearms, power plants, etc. 

Some pioneering studies in the late twentieth century \cite{Spudich1972,Spudich1994,Alberts}  led to the discovery of the so-called molecular motors that carry out innumerable functions at the intra and inter-cellular scales within living organisms. They have often been studied using the popular model of Brownian ratchets \cite{Astumian2001,Astumian2002,Ait-Haddou2003}, which was earlier introduced by Feynman to explain the role of thermal fluctuations in small systems \cite{Feynman1986}.
This shows that it is not only possible to have microscopic versions of engines, but also to run them efficiently even when the dynamics is significantly affected by thermal fluctuations. Several such engines have been realized both theoretically \cite{sei08_epl,Rana2014,Abah2014,Verley2014} and experimentally \cite{bechinger2012,Koski2014,Abah2016,Goold2019}. For an excellent review on classical microscopic heat engines, see \cite{roldan2016}. Devising of autonomous stochastic engines has also been studied \cite{Garcia2016,sei19_prx}. 
Several potential uses of these tiny devices have been put forward in the literature \cite{Saadeh2014}. 

One of the several ramifications of the area of stochastic heat engines is the preparation of small engines using active fluctuations, pioneered by the experimental work of \cite{ajay2016_nature}. Theoretical studies of active heat engines have been carried out in \cite{Saha_2019,lahiri2020,Kumari2021}. These theoretical works primarily focused on heat engines that used either single active particle as the working system in presence of a thermal heat bath, or a single passive particle in contact with an active heat bath. 
If our working system consists of multiple active particles that interact with each other, the situation changes drastically, since the dynamics of a collection of interacting particles is statistically different from an ensemble of single particles. For instance, if the particles are self-propelled, then the collection often shows collective motion \cite{Rana2019}, frequently observed in the patterns of movements of a flock of birds, school of fishes, etc. Given the changes in the statistics of motion, it is natural to ask whether such a set of particles can give rise to qualitative changes in the thermodynamic properties of small heat engines.  
We address this question by studying a collection of self-propelled particles undergoing polar alignment interaction. 
We confine a collection of such active particles in a circular trap subjected to a time-dependent wall steepness to set up a multiparticle stochastic heat engine.

\section{The Model}

Consider a system containing $N$ active particles that interact with each other in a Vicsek-like manner \cite{Vicsek1995}. Each particle observes the direction of motion of its neighbourhood (defined as those that occur within a circle of a given radius $l_0$ with the concerned particle at its centre), and attempts to align its own direction of motion with the average of those of its neighbours. We denote this average direction by $\langle \hat{n}\rangle_{l_0}$. In general, this alignment is not perfect. This imperfection is introduced by means of a rotation matrix $R_\alpha(\theta)$ that randomly chooses a ``rotated'' direction given by the angle of deviation $\theta$ with the computed average direction $\langle \hat{n}\rangle_{l_0}$, such that $\theta \in (0,\alpha)$.  This dynamics is incorporated by defining the following active force of strength $f_0$ acting on the $i^{\rm th}$ particle, as given below:
\begin{align}
    {\bf F}_i^{\rm act} \equiv f_0 R_\alpha(\theta)\cdot \langle \hat{n}\rangle_{l_0}.
    \label{eq:ActiveForce}
\end{align}
 While simulating, in order to find the average direction, we first compute the total velocity vectors in the $x$ and $y$ directions, given by $\textbf{V}_{x,i\in l_0} \equiv \sum_{i\in l_0}\textbf{v}_{x,i}$ and $\textbf{V}_{y,i\in l_0} \equiv \sum_{i\in l_0}\textbf{v}_{y,i}$ respectively. Subsequently, the average direction in the neighbourhood of the $i^{\rm th}$ particle is given by
\begin{align}
    \hat{\textbf{n}}_x = \frac{\textbf{V}_{x,i\in l_0}}{\sqrt{\textbf{V}_{x,i\in l_0}^2 + \textbf{V}_{y,i\in l_0}^2}}; \hspace{1cm} \hat{\textbf{n}}_y = \frac{\textbf{V}_{y,i\in l_0}}{\sqrt{\textbf{V}_{x,i\in l_0}^2 + \textbf{V}_{y,i\in l_0}^2}}.
\end{align}
In addition to the active force,  an inter-particle repulsive force is derived from Weeks-Chandler-Andersen (WCA)  interaction of strength $\epsilon$ \cite{Weeks1971}. If the distance between particles $i$ and $j$ is given by $r_{ij}$, while the effective radius of each particle is $\sigma$, then the WCA potential is given by,
\begin{alignat}{2}
    V_{ij}^{\rm WCA} &= 4\epsilon\left(\frac{ \sigma^{12} }{ r_{ij}^{12} } - \frac{ \sigma^{6} }{ r_{ij}^{6}  }\right) + \epsilon, \hspace{2cm} && r_{ij} < 2^{1/6}\sigma; \nonumber\\
    & =0 && \mbox{otherwise}.
    \label{eq:WCA_potential}
\end{alignat}
\begin{figure}[!ht]
    \centering
   
   \begin{subfigure}{0.48\linewidth}
       \includegraphics[width=\linewidth]{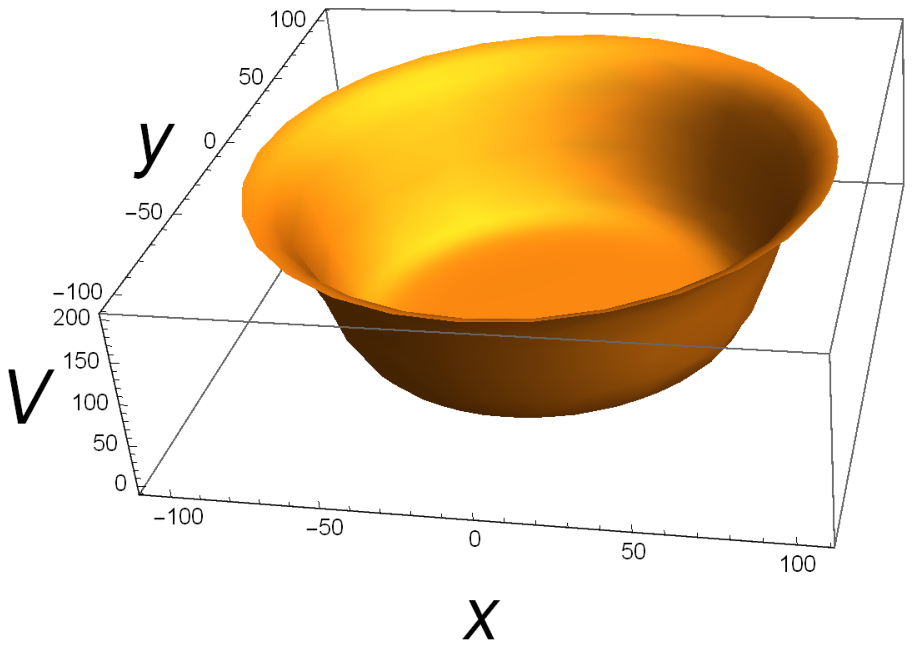}
       \caption{}
   \end{subfigure}
    \begin{subfigure}{0.48\linewidth}
       \includegraphics[width=\linewidth]{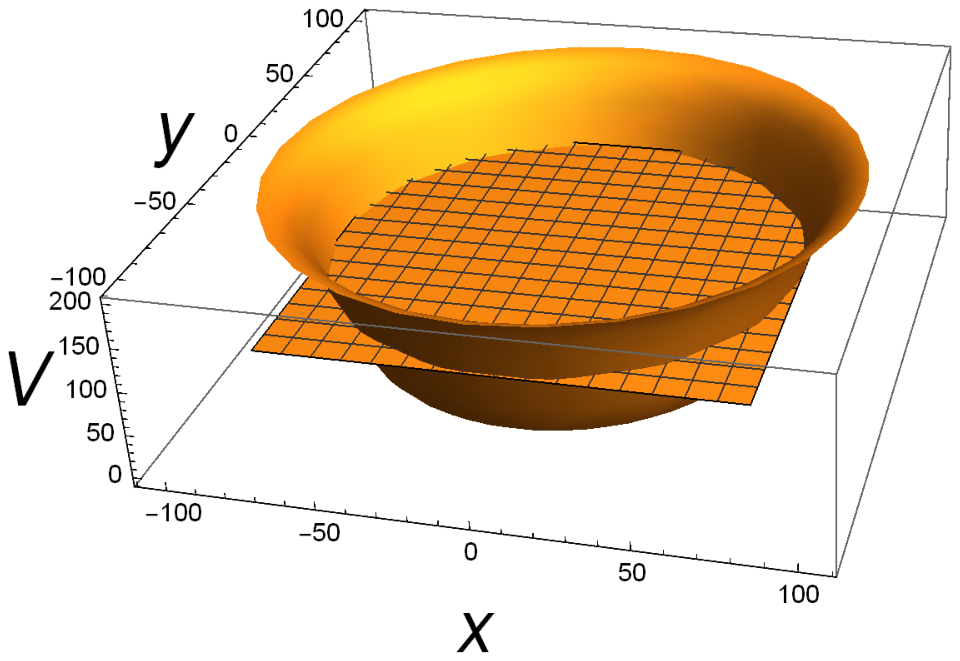}
       \caption{}
   \end{subfigure}
    \caption{(a) Potential described by Eq. \eqref{eq:potential}, with parameters chosen as per convenience to enhance clarity. (b) Same potential with the ceiling introduced at $V=V_0$.}
    \label{fig:potential}
   \end{figure}

\begin{figure}[!h]
    \centering
    \includegraphics[width=0.5\linewidth]{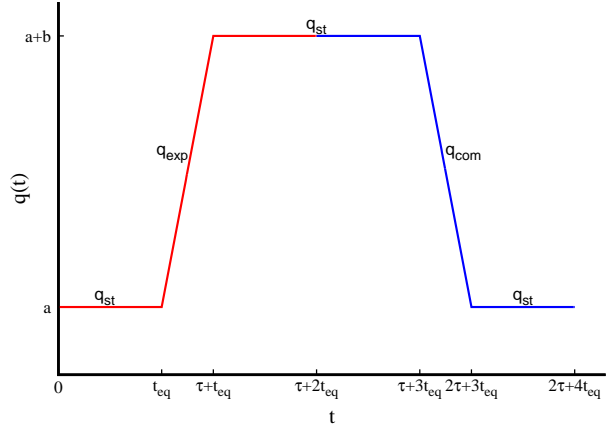}
    \caption{Plot showing the time variation of stiffness parameter. The red and blue portions correspond to higher and lower values of $f_0$, respectively.}
    \label{fig:stiff}
\end{figure}

The particles are trapped in a potential given by (see Fig. \ref{fig:potential}(a))
\begin{align}
    V(r_i,t) &= V_0 [1 + \tanh\{q(t)(r_i-r_0)\}],
    \label{eq:potential}
\end{align}
where $r_i \equiv |\textbf{r}_i|$, with $\textbf{r}_i$ being the position vector of the $i^{\rm th}$ particle. The height of the potential is $2V_0$, which is chosen to be sufficiently high so that none of the particles is allowed to escape. $q(t)$ is a measure of the steepness of the walls of the potential, and is made time-dependent to allow the potential to change its shape cyclically so as to resemble the driving of a Stirling engine \cite{bechinger2012}. For convenience, we choose this function to be linear in time (see fig. \ref{fig:stiff}):
\begin{alignat}{2}
    q_{\rm st} &= a, \hspace{5cm}&& 0<t\le t_{\rm eq} \nonumber\\
    q_{\rm exp}(t) &= a + b \left(\frac{t-t_{\rm eq}}{\tau}\right),  && t_{\rm eq}<t\le t_{\rm eq}+\tau \nonumber\\
    q_{\rm st} &= a+b, && \tau+t_{\rm eq}<t\le \tau+3t_{\rm eq} \nonumber\\
    q_{\rm com}(t) &= a + b \left(1-\frac{t-[\tau+2t_{\rm eq}]}{\tau} \right), && \tau + 3t_{\rm eq} <t\le 2\tau + 3t_{\rm eq} \nonumber\\
    q_{\rm st} &= a, && 2\tau+3t_{\rm eq} < t\le 2\tau+4t_{\rm eq}.
    \label{eq:com}
\end{alignat}
Here, $a$ and $b$ are two constants. The subscripts ``${\rm exp}$'' and ``${\rm com}$'' correspond to the expansion and the compression strokes of the engine cycle, each being of duration $\tau$. The temperature of the bath is kept constant throughout the cycle (single heat bath), but the activity $f_0$ of the system is changed periodically with $q(t)$. The value of activity is  is higher, ($f_0=f_0^e$) during the time interval $0<t\le \tau+2t_{\rm eq}$, and lower ($f_0=f_0^c$) during the time interval $\tau + 2t_{\rm eq} <t\le 2\tau + 4t_{\rm eq}$ (see Fig. \ref{fig:stiff}). 
Note that during the expansion stroke, the steepness parameter $q_{\rm exp}(t)$ changes linearly from $a$ to $a+b$. In the compression stroke, as per Eq. \eqref{eq:com}, the value of $q_{\rm com}(t)$ reverts from $a+b$ to $a$ linearly with time. At both junctions between the expansion and compression arms, the system is allowed to relax to equilibrium. 

As shown in Fig. \ref{fig:potential}(b), a ceiling is introduced at the $V=V_0$ plane of the trap so as to constrain the particles to stay below it, which ensures that the volume to which they are confined increases or decreases when the steepness decreases or increases, respectively (note that in absence of the ceiling, the volume of the trap becomes independent of $q(t)$).  

It has been shown earlier in \cite{Rana2019} that by changing the steepness of such a trap, one can tune the self-organisation of the active particles within the trap.  In particular, if the steepness is high enough the particles form a disordered layer attached to the boundary whereas on lowering the steepness below a certain threshold, they form hexagonally ordered round clusters that simultaneously rotate along the boundary and roll around its centre of mass. 
In our case, we work with a much smaller number of particles, namely $5\%$ of \cite{Rana2019}. Consequently, we find that even though the activity leads to the clustering of these particles, and the steepness of the trap determines the proximity of the cluster to the wall, the above mentioned order-to-disorder transition is absent. In this work, we study the thermodynamics of such a system forming an engine by means of periodic variations in the trap steepness.

The equation of motion of the system of particles is given by the underdamped Langevin equation:
\begin{align}
    m\frac{d{\bf v}_i}{dt} &= -\gamma {\bf v}_i + {\bf F}_i^{\rm pot} + {\bf F}_i^{\rm act} + (\sqrt{2\gamma k_B T})\bm{\xi}_i,
    \label{eq:EoM}
\end{align}
where
\begin{align}
    {\bf F}_i^{\rm pot} &= -\nabla_i V - \sum_j\nabla_{ij} V_{ij}^{\rm WCA}.
\end{align}
Here, $T$ is the temperature of the heat bath, and $\gamma$ is the friction coefficient. The Boltzmann constant $k_B$ is set to unity throughout this work for convenience.
The notation $\nabla_i$ consists of derivatives with respect to the particle positions $r_i$, whereas $\nabla_{ij}$ is the derivative with respect to the interparticle distance $r_{ij}$. The noise $\bm{\xi}_i$ acting on the $i^{\rm th}$ particle is sampled from a standard normal distribution, and has the properties: $\langle \xi_i(t)\xi_j(t')\rangle = \delta_{ij}\delta(t-t')$. 

\section{Thermodynamic observables}

The system being of microscopic dimensions, it will be described by the trajectory-dependent thermodynamic quantities as defined in stochastic thermodynamics \cite{sek98,sekimoto,sei12_rpp}.
 The various thermodynamic observables that we would address in this work are defined below. 
\paragraph{{\bf Work:}} The extracted work/power constitutes one of the most important attributes of any engine. The stochastic work $W_i$ done \emph{on} the $i^{\rm th}$ particle and the particle-averaged work $W$ done on the system are defined by:
\begin{align}
W_i(t) &= \int_0^t \frac{\partial V}{\partial t}dt = V_0 \int_0^t dt \dot{q}(t)~\mbox{sech}^2\{q(t)(r_i-r_0)\},\nonumber\\
W(t) &= \frac{1}{N}\sum_{i=1}^N W_i(t).
\label{eq:work}
\end{align}
Here, the overhead dot denotes a time derivative. 

\paragraph{{\bf Internal energy change:}}  Performing Stratonovich product \cite{ris} (denoted by the symbol $\circ$) on both sides of Eq. \eqref{eq:EoM} by $\textbf{dr}_i = \textbf{v}_i dt$ and rearranging terms, we get
\begin{align}
    \big(\gamma \textbf{v}_i - \sqrt{2\gamma k_B T}~\bm{\xi}_i\big)\circ \textbf{v}_i dt = -\frac{d}{dt}\left(\frac{1}{2}m\textbf{v}_i^2\right) - \nabla_i V(r_i,t)\circ \textbf{dr}_i + (\textbf{F}_i^\text{WCA} + \textbf{F}_i^\text{act})\circ \textbf{v}_i dt.
\end{align}
Here, $\textbf{F}_i^\text{WCA}\equiv -\sum_j\nabla_{ij}V_{ij}^\text{WCA}$. The left hand side is identified with infinitesimal heat dissipated ($dQ_i$) into the bath by the $i^{\rm th}$ particle in time $dt$. 
In the right hand side, using chain rule, we obtain $\nabla_i V(r_i,t) \circ \textbf{dr}_i = dV(\textbf{r}_i) - \frac{\partial V}{\partial t}dt$. We can then rewrite the above equation as
\begin{align}
    dQ_i = dW_i - dE_i,
    \label{eq:HeatDissipated}
\end{align}
where 
\begin{align}
    dE_i = d\left(\frac{1}{2}m\textbf{v}_i^2 + V(r_i,t)\right) - (\textbf{F}_i^\text{WCA} + \textbf{F}_i^\text{act})\circ \textbf{v}_i dt.
    \label{eq:InternalEnergyChange}
\end{align}
Note that here, as the particles are interacting among themselves via the WCA potential and the active force, they both contribute to the internal energy of  the particle and subsequently to the particle-averaged internal energy, $\Delta E \equiv \frac{1}{N}\sum_i dE_i$, of the whole system. 
Integrating both sides of Eq. \eqref{eq:HeatDissipated} over time, we obtain the net change in the dissipated heat, work done and internal energy over a trajectory. Finally, averaging over the particles will yield the first law in terms of the particle-averaged quantities $Q$, $W$ and $\Delta E$, given by 
\begin{align}
    Q = W - \Delta E.
    \label{eq:FirstLaw}
\end{align}
Using Eq. \eqref{eq:FirstLaw}, we can compute the value of $Q$.
Note, however, that the internal energy $E$ is \textit{not a state function} in the above case, but depends on the trajectory taken by the concerned particle. This is in stark contrast to the internal energy of AOUP particles \cite{lahiri2020,Saha_2019}.  
This is due to the fact that the change in energy caused by the active forces has a contribution from energy exchanges with hidden degrees of freedom that are not observed in the experiment \cite{Barato2022}. 
The direct application of the concepts of stochastic thermodynamics \cite{sekimoto} to active systems can give rise to such unexpected results, and as yet a consensus regarding these definitions is lacking \cite{Fodor2022}.

\section{Results and Discussions}

\subsection{Different configurations of the system of particles}

\begin{figure}[h!]
    \centering
    \begin{subfigure}{0.3\linewidth}
        \includegraphics[width=\linewidth]{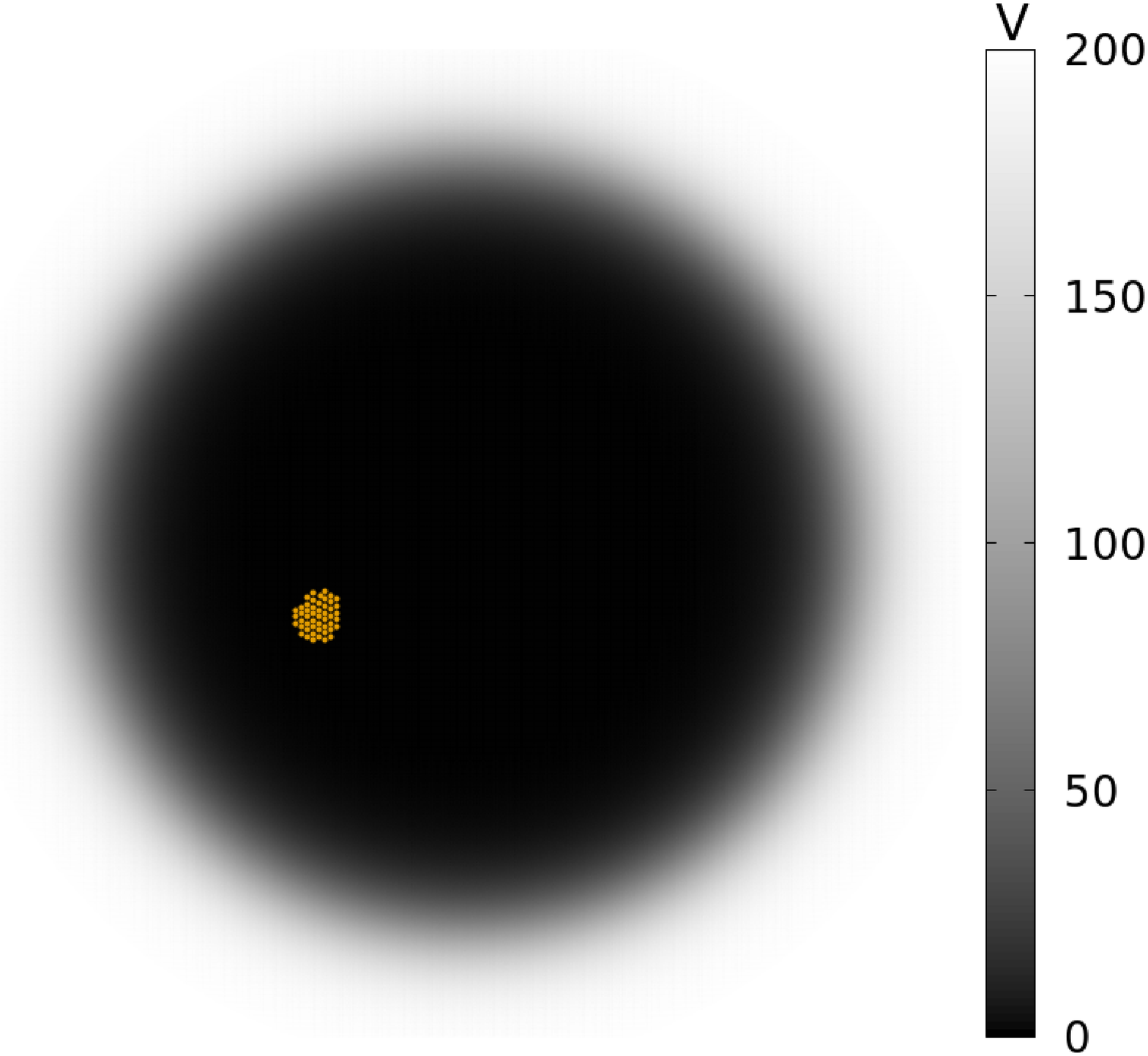}
        \caption{$t=t_{\rm eq}$}
    \end{subfigure}
    \begin{subfigure}{0.3\linewidth}
        \includegraphics[width=\linewidth]{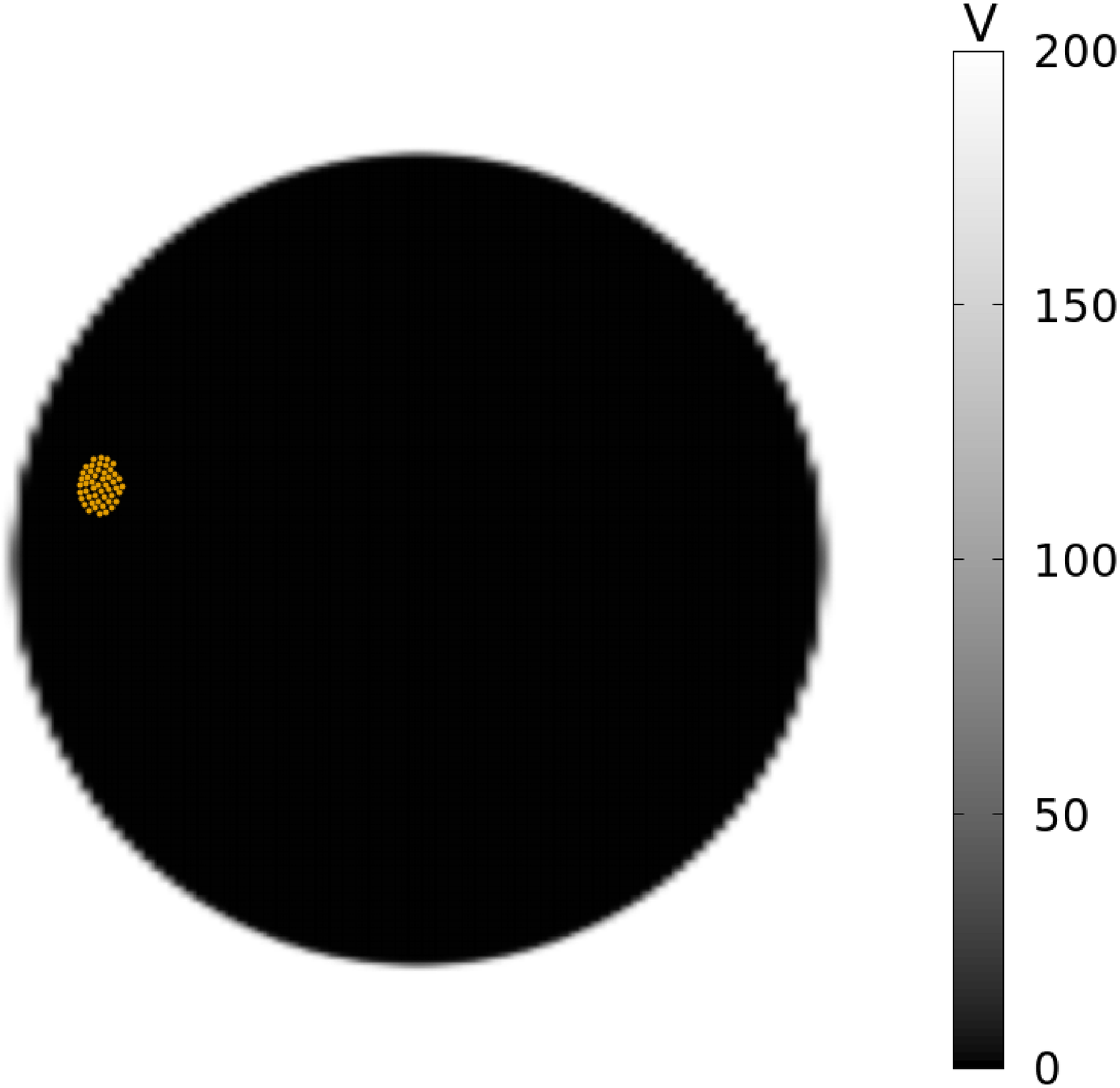}
        \caption{$t=t_{\rm eq}+\tau$}
    \end{subfigure}
     \begin{subfigure}{0.3\linewidth}
        \includegraphics[width=\linewidth]{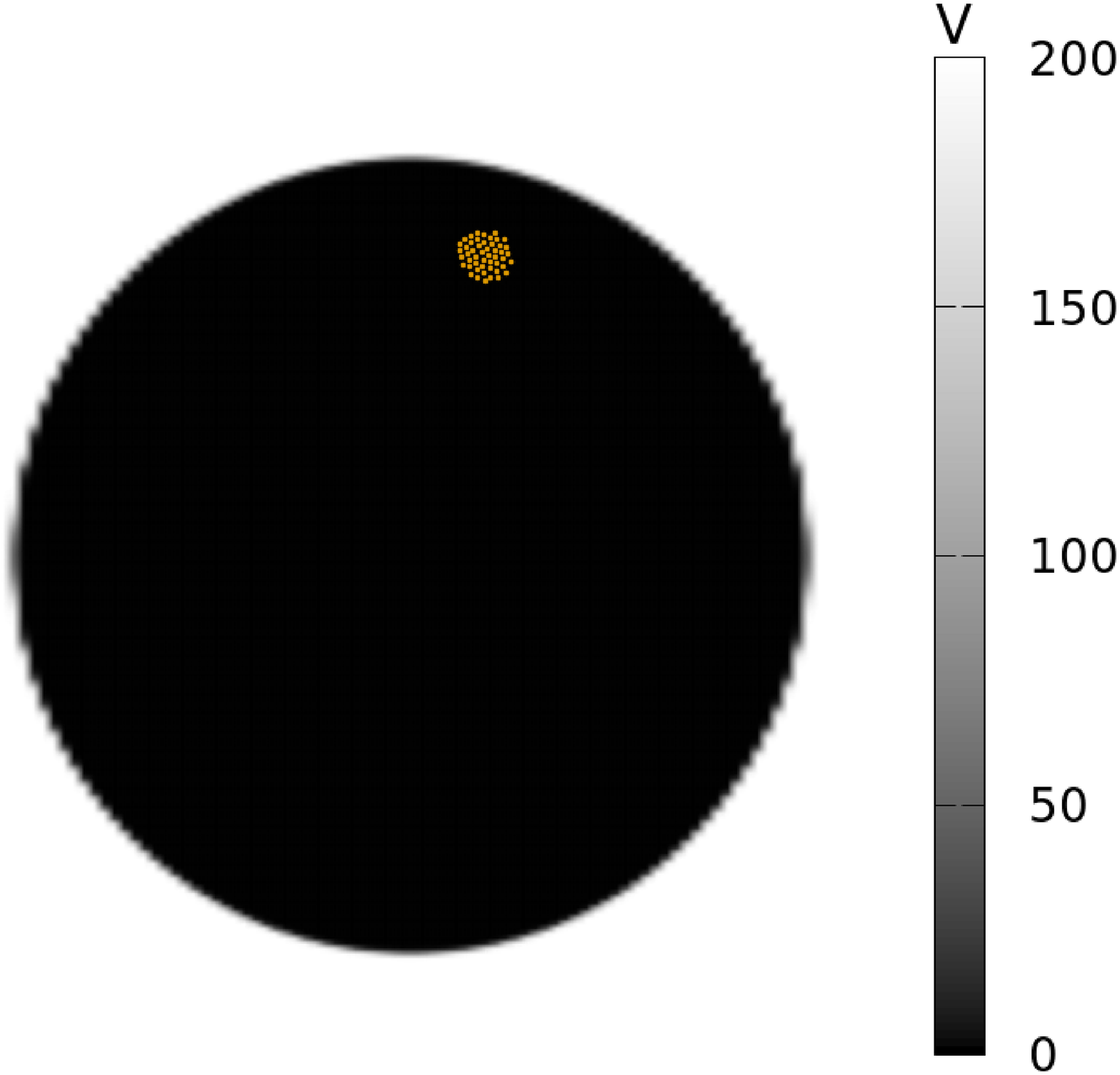}
        \caption{$t=3t_{\rm eq}+\tau$}
    \end{subfigure}
     \begin{subfigure}{0.3\linewidth}
        \includegraphics[width=\linewidth]{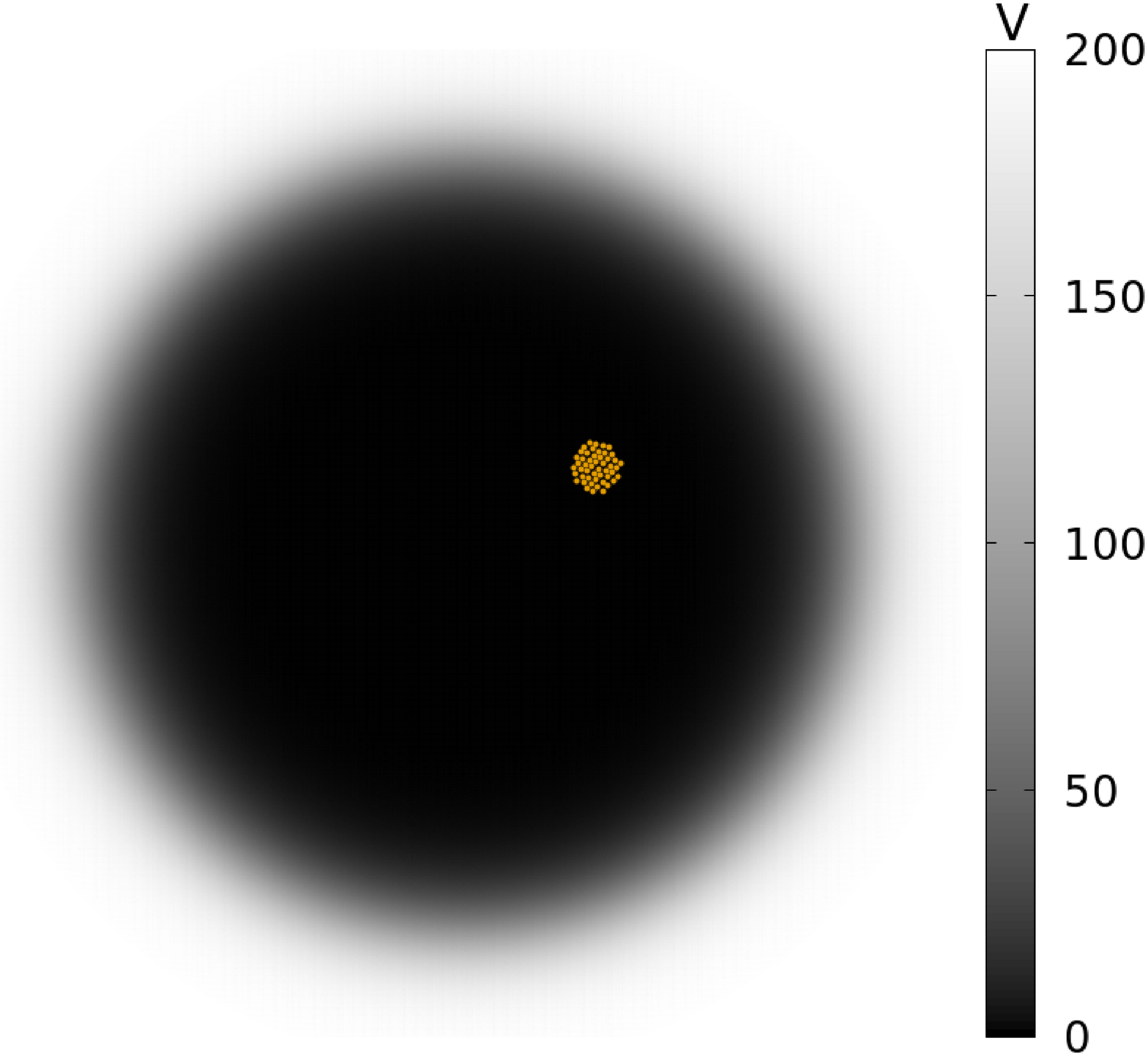}
        \caption{$t=3t_{\rm eq}+2\tau$}
    \end{subfigure}
    \begin{subfigure}{0.3\linewidth}
        \includegraphics[width=\linewidth]{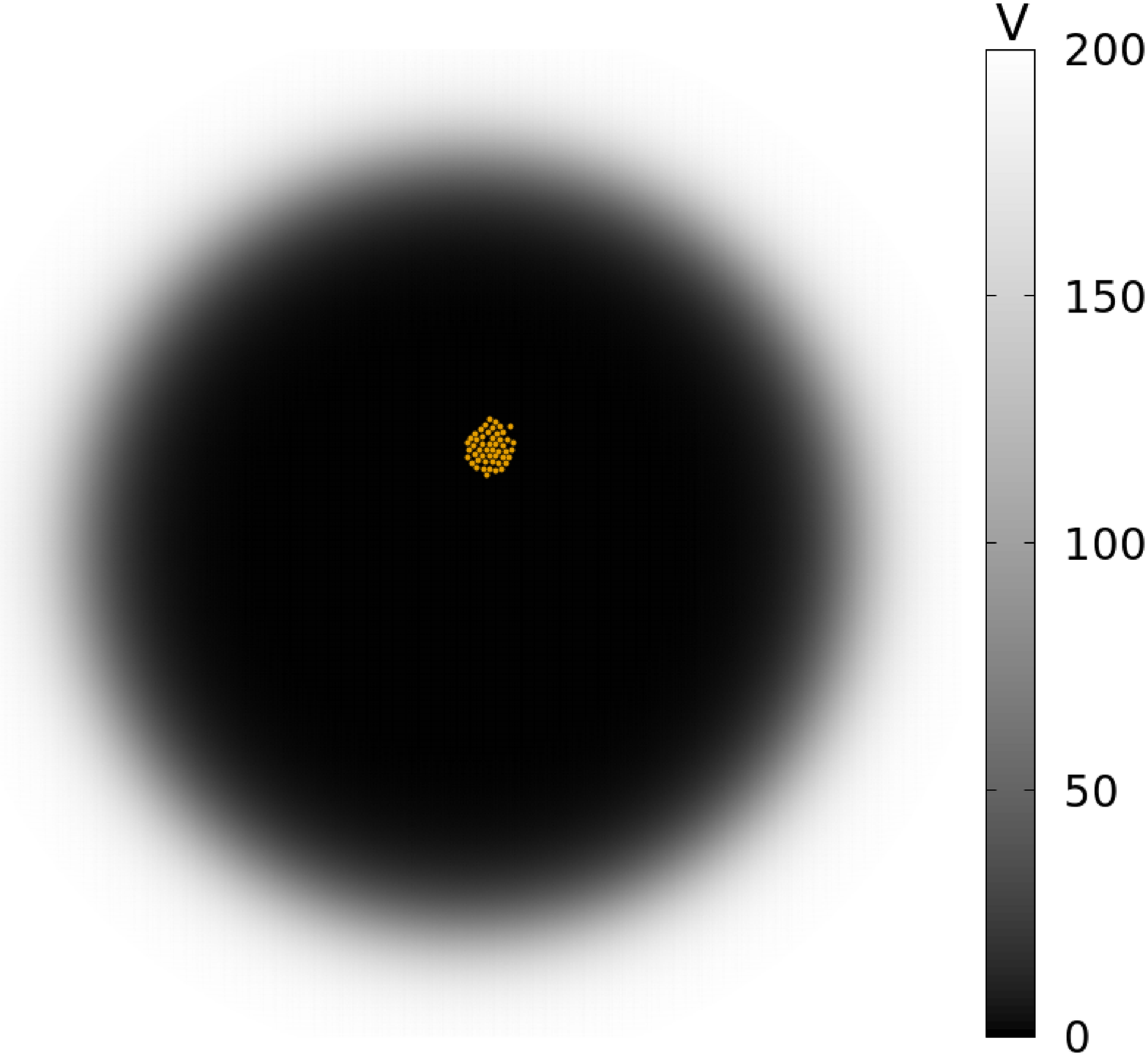}
        \caption{$t=4t_{\rm eq}+2\tau$}
    \end{subfigure}
    
    \caption{
    The plots show particle positions by orange dots and the potential in gray scale.
   The trap along with the particle positions at (a) $t=t_{\rm eq}$ (b) $t=t_{\rm eq}+\tau$ (c) $t=3t_{\rm eq}+\tau$ (d) $t=3t_{\rm eq}+2\tau$ (e) $4t_{\rm eq}+2\tau$. Other parameters are: $r_0=2,~V_0=100,~\gamma=1,~l_0=2,~T_c=T_h=0.001,~f_{0}^e=20,~f_{0}^c=10$.}

    \label{fig:phases}
\end{figure}

We show in Fig. \ref{fig:phases} the various configurations that the system of particles attains during the course of a single engine cycle in a time-periodic steady state.  The initial configuration, shown in Fig. \ref{fig:phases}(a), consists of the particles forming a round-shaped cluster, far from the circular wall (with radius $r_0$ steepness $a$). 
Next, we begin the protocol given by Eq. \eqref{eq:com}, where the value of $q(t)$ is made to increase  from $a$ to $a+b$. The sharpness of the wall of the potential is clear from the well-defined circumference of the disk in figure \ref{fig:phases}(b). The cluster is now observed to have moved close to the wall of the trap.
At this stage, we keep the value of the steepness fixed at $a+b$ and allow a relaxation for time $2t_{\rm eq}$ (figure \ref{fig:phases}(c)). 
 Fig. \ref{fig:phases}(d) shows the distribution of the particles after evolving via Eq.  \eqref{eq:com} from $t=3t_{\rm eq}+\tau$ to $t=3t_{\rm eq}+2\tau$, during which the steepness parameter returns from $a+b$ to $a$. In this case, the cluster goes far from the wall. 
Fig. \ref{fig:phases}(e) gives the distribution after the system has undergone a final relaxation, with the steepness parameter fixed at the value $a$. The cluster is observed to remain far from the wall during this time. This completes the entire cycle.

\subsection{Study of thermodynamics of the engine}

There are a multitude of parameters that may affect the thermodynamic behaviour of the multiparticle heat engine under study. They are the activity strength $f_0$ (see Eq. \eqref{eq:EoM}), the trap radius $r_0$, the angular noise $\alpha$, the trap height $V_0$, and the number density of the particles. We choose to confine our investigation to the dependence of the engine output on the parameters $f_0,~\alpha$ and $l_0$. The active force strengths in the expansion and compression halves will be denoted by $f_{0}^e$ and $f_{0}^c$, respectively.

\begin{figure}[!ht]
    \centering
   
   \begin{subfigure}{0.48\linewidth}
       \includegraphics[width=\linewidth]{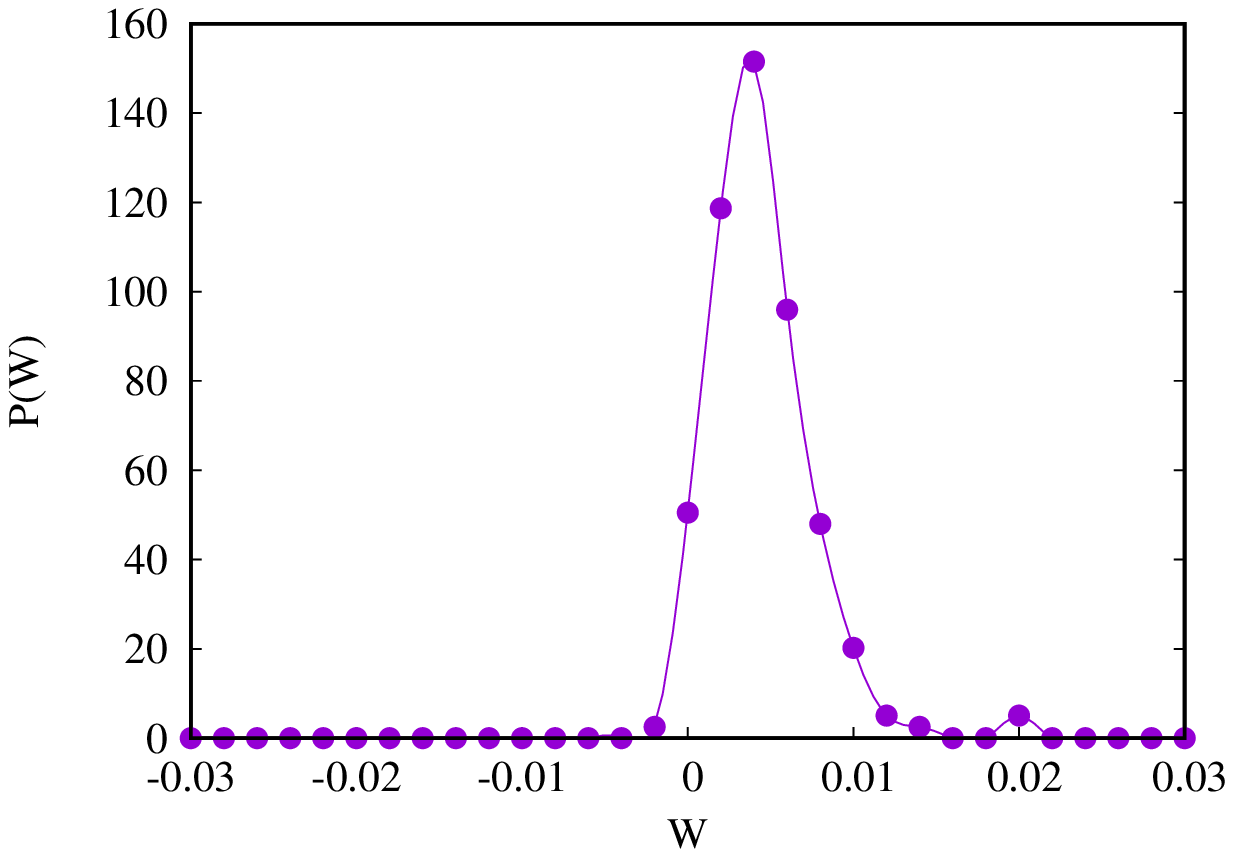}
       \caption{}
   \end{subfigure}
    \begin{subfigure}{0.48\linewidth}
       \includegraphics[width=\linewidth]{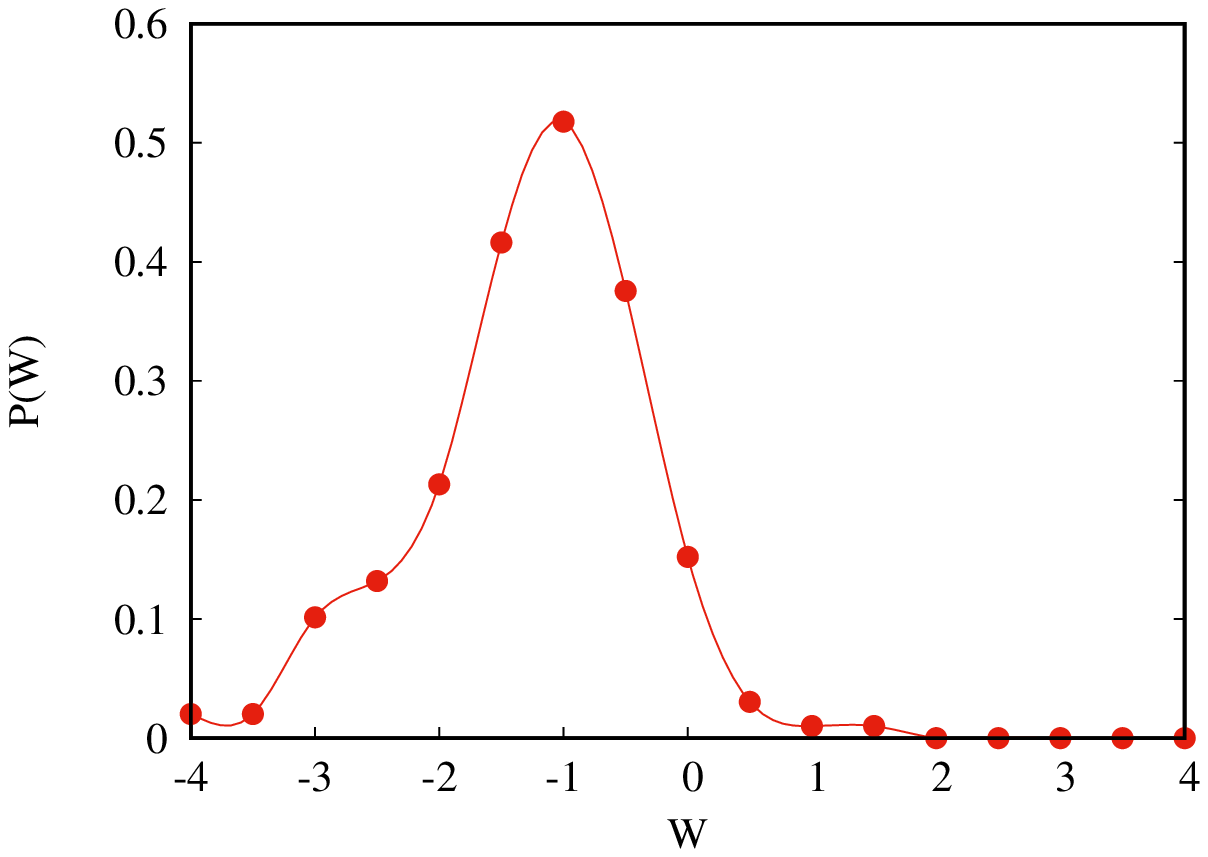}
       \caption{}
   \end{subfigure}

    \caption{(a) Plot showing the distribution of work done over an engine cycle without activity.  (b) Plot showing the distribution of work done over an engine cycle with activity. Parameters used: $\alpha=1$, $f_{0}^e=20$, $f_{0}^c=10,~T_h=T_c=0.001$.}
    \label{fig:PW}
\end{figure}

\paragraph{Dependence of work on the strength of activity:}
In Fig. \ref{fig:PW}(a), the distribution of work done \emph{on} the system over a full engine cycle in steady state in absence of activity ($f_0^e = f_0^c = 0$) has been shown. The system is made to interact with a single thermal reservoir during the cycle. The other parameters are as mentioned in the caption. Note that in our convention, work \emph{extracted} is negative. 
The mean work is observed to have a small positive value (work is done on the system). The mean work being positive (instead of zero) indicates the nonequilibrium nature of the process. 
Figure \ref{fig:PW}(b), on the other hand, clearly shows that in presence of finite activity when the system undergoes expansion and compression, given by $f_{0}^e=20$ and $f_{0}^c=10$ respectively, work can indeed be extracted even in the presence of a single heat bath.
This crucial property is a major distinction from heat engines using passive particles, where at least two thermal baths are required to get a work output. A similar property, even for an engine using a single active particle, was demonstrated theoretically in \cite{Saha_2019}.

\begin{figure}[!h]
    \centering
    \includegraphics[width=0.5\linewidth]{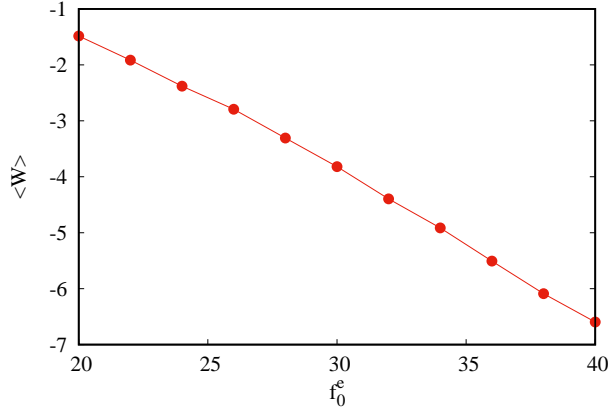}
    \caption{Plot showing the variation of mean work with expansion cycle activity ($f_{0}^e$). The parameters are: $N=50$, $l_0=2$, $f_{0}^c=10,~T_h=T_c=0.001$, $\alpha=1$}
    \label{fig:W_activity}
\end{figure}

Fig. \ref{fig:W_activity} shows the variation of average work with $f_0^e$, keeping $f_0^c$ constant. We find that the extracted work increases with increase in $f_0^e$.  Values of the different parameters are as mentioned in the figure caption. It is evident that work can indeed be extracted from the engine in connection with a single bath but with two different activities in the two halves of the cycle. Such an effect was also observed in \cite{Saha_2019}, where single active Ornstein-Uhlenbeck particles were used. 
This is due to the fact that during the expansion cycle, the particles being more active, tend to be in contact with the boundary of the trap for a substantial amount of time (as can be observed from the snapshots shown in figure \ref{fig:phases}). Since the potential is higher closer to the boundary, this tendency leads to a larger extraction of work. In the compression half, the same effect gives rise to pumping of work into the system, but to a much lesser degree owing to the reduced strength in activity.

\begin{figure}[h!]
    \centering
    \begin{subfigure}{0.43\textwidth}
        \includegraphics[width=\textwidth]{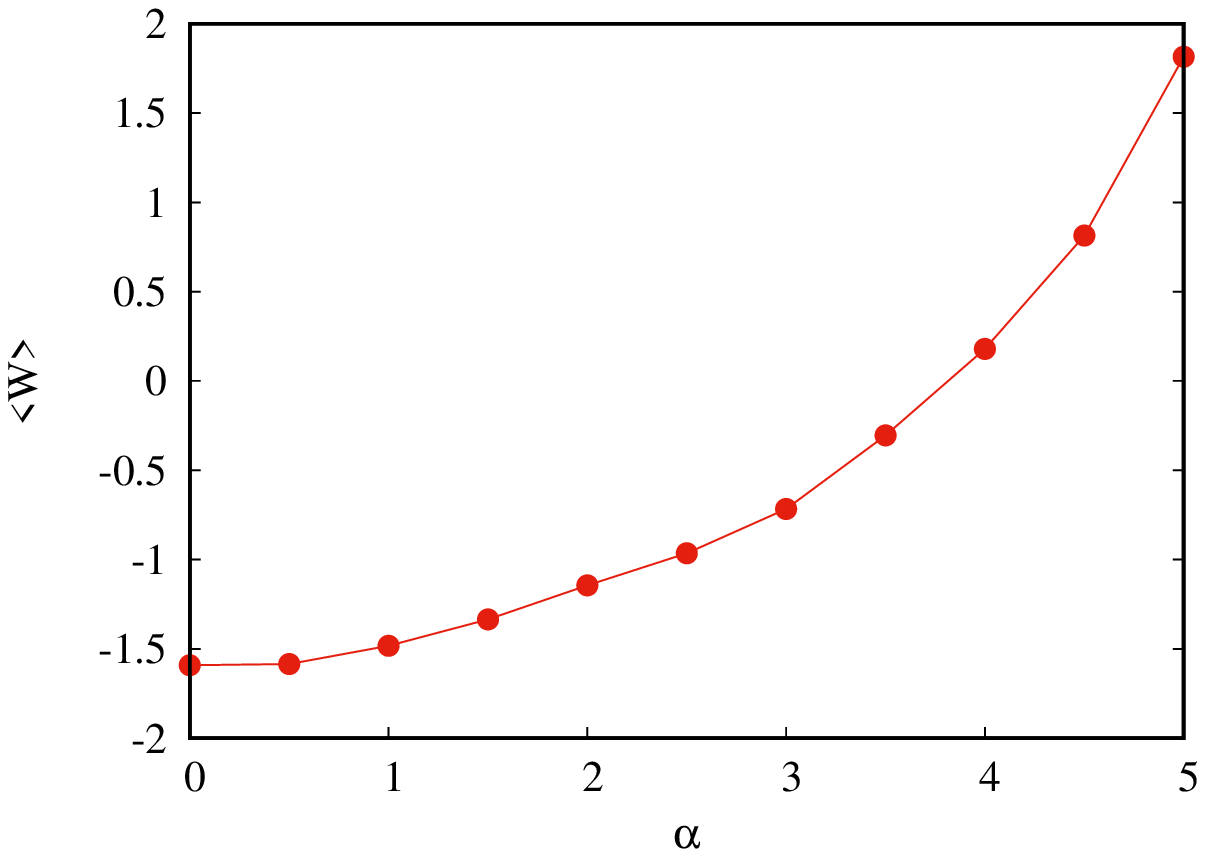}
        \caption{}
    \end{subfigure}
    \begin{subfigure}{0.43\textwidth}
         \includegraphics[width=\textwidth]{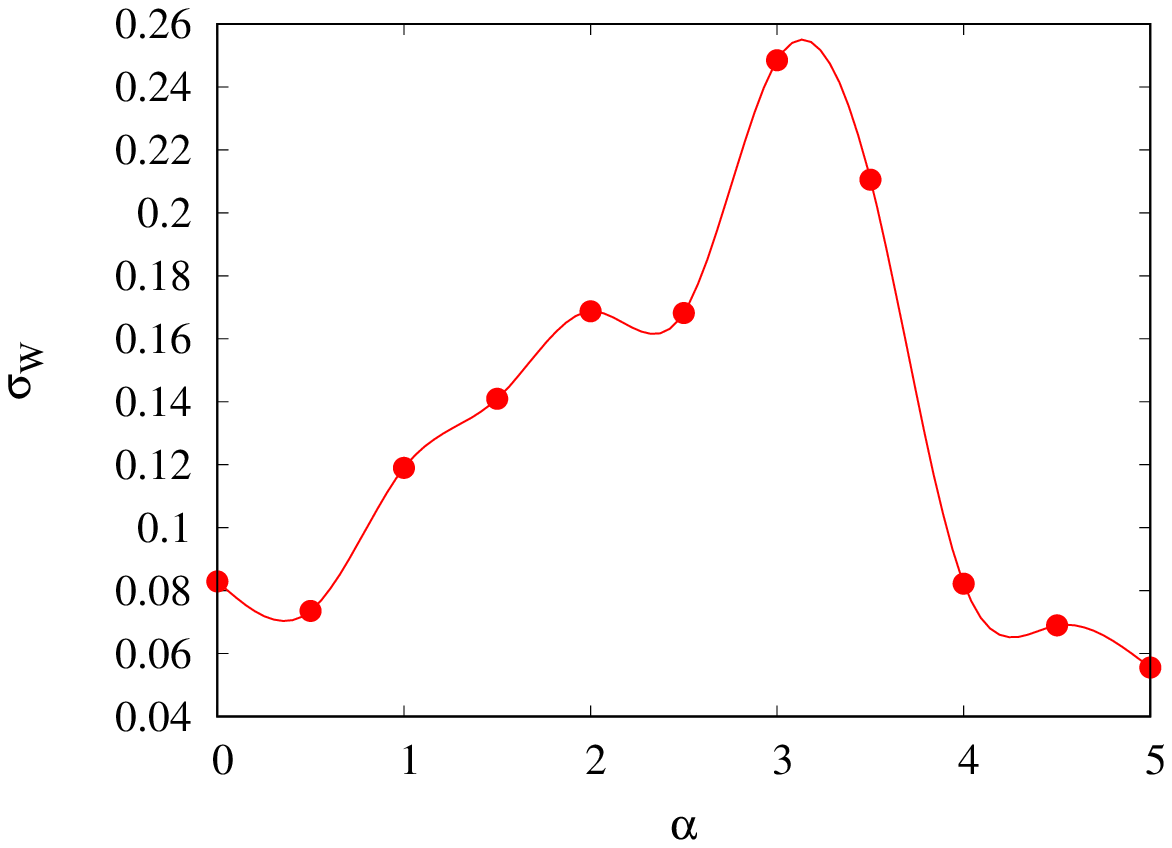}
         \caption{}
    \end{subfigure}
    \centering
    \begin{subfigure}{0.43\textwidth}
        \includegraphics[width=\textwidth]{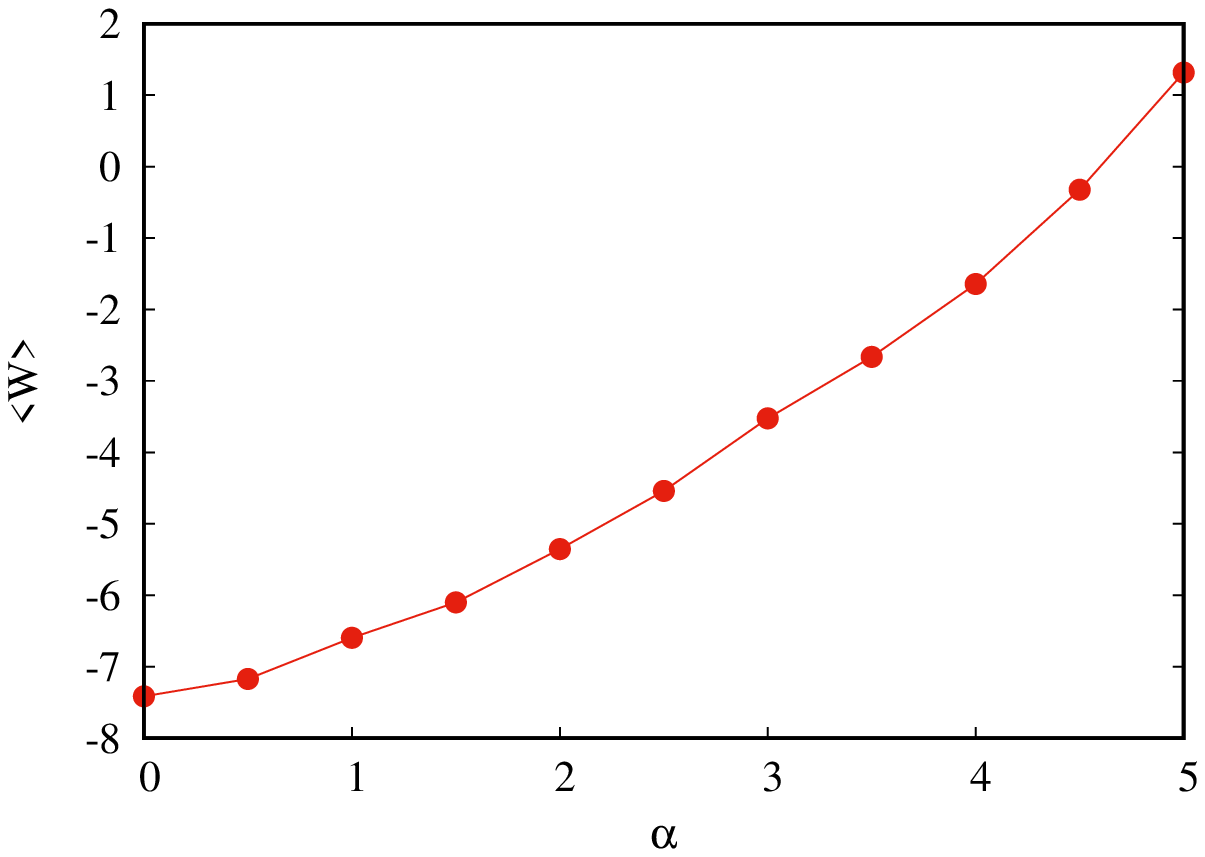}
        \caption{}
    \end{subfigure}
       \begin{subfigure}{0.43\textwidth}
         \includegraphics[width=\textwidth]{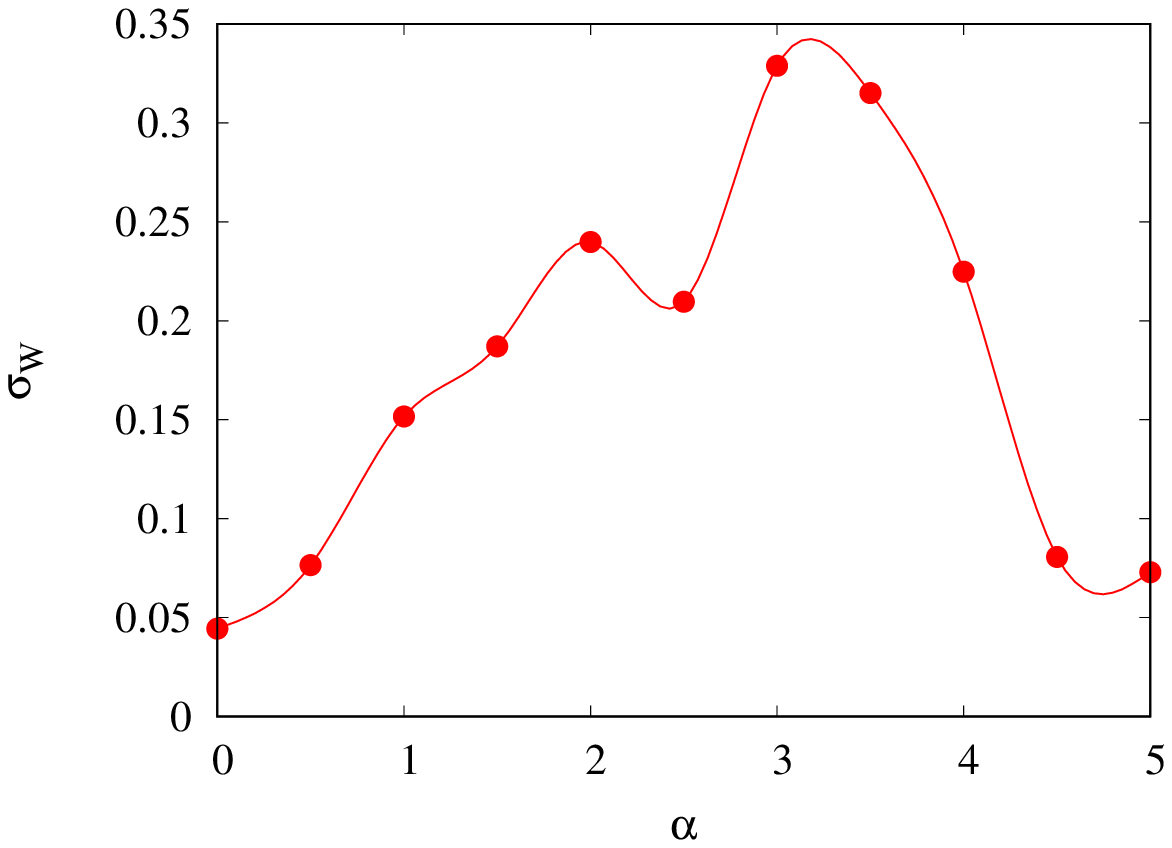}
         \caption{}
    \end{subfigure}
    \caption{(a) Plot showing the variation of mean work with $\alpha$. The parameters are: $f_{0}^e=20$, $f_{0}^c=10,~T_h=T_c=0.001$, $l_0=2$. (b) Plot showing variance in work as a function of $\alpha$ for $f_{0}^e=20$. (c) Same plot as (a), with $f_{0}^e=40$. (d) Same plot as (b), with $f_{0}^e=40$.}
    \label{fig:W_eta}
\end{figure}

\paragraph{Dependence of work and its variance on angular noise:}
We next turn our attention to the dependence of the extracted work on the angular noise $\alpha$. 
In Fig. \ref{fig:W_eta}(a) and (c) we have shown the variation of average extracted work with the angular noise $\alpha$. The activity in the expansion stroke given by $f_{0}^e=20$ and 40, respectively (for the values of other parameters, see figure caption). We find a significant variation in average work with increase in $\alpha$, and eventually the system crosses over into the refrigerator mode (the mean work is done on the system, while heat is dissipated on average into the bath during expansion)  beyond $\alpha=4$. 
Physically, this means that the particles that can  align with their neighbours with higher accuracy typically outperform the ones exhibiting more randomness in the alignment. Furthermore, it is evident that by tuning the angular noise, one can switch between engine and refrigerator modes.

The variances in extracted work is a measure of the reliability of the engine's performance. Keeping this in view, we compute the variances of work (corresponding to $f_0^e=20$ and 40) for different values of $\alpha$ in Figs. \ref{fig:W_eta}(b) and (d). The variance is found to exhibit non-monotonicity as a function of the angular noise. It shows a peak at $\alpha\approx \pi$, where the reliability of the extracted work will be lower. Consequently, these values may be avoided in order to enhance reliability over an engine cycle. Furthermore, the reliability of the output decreases with the increase in the active force strength $f_{0}^e$, as observed from the increase in the values of the variance. This means at very high activity of the constituent particles of the working system, the dispersion of work from one cycle to another is greater, even though the mean extracted work is more. This implies a trade-off between the mean work and dispersion, with activity.

\begin{figure}[h!]
    \centering
    \begin{subfigure}{0.45\textwidth}
        \includegraphics[width=\textwidth]{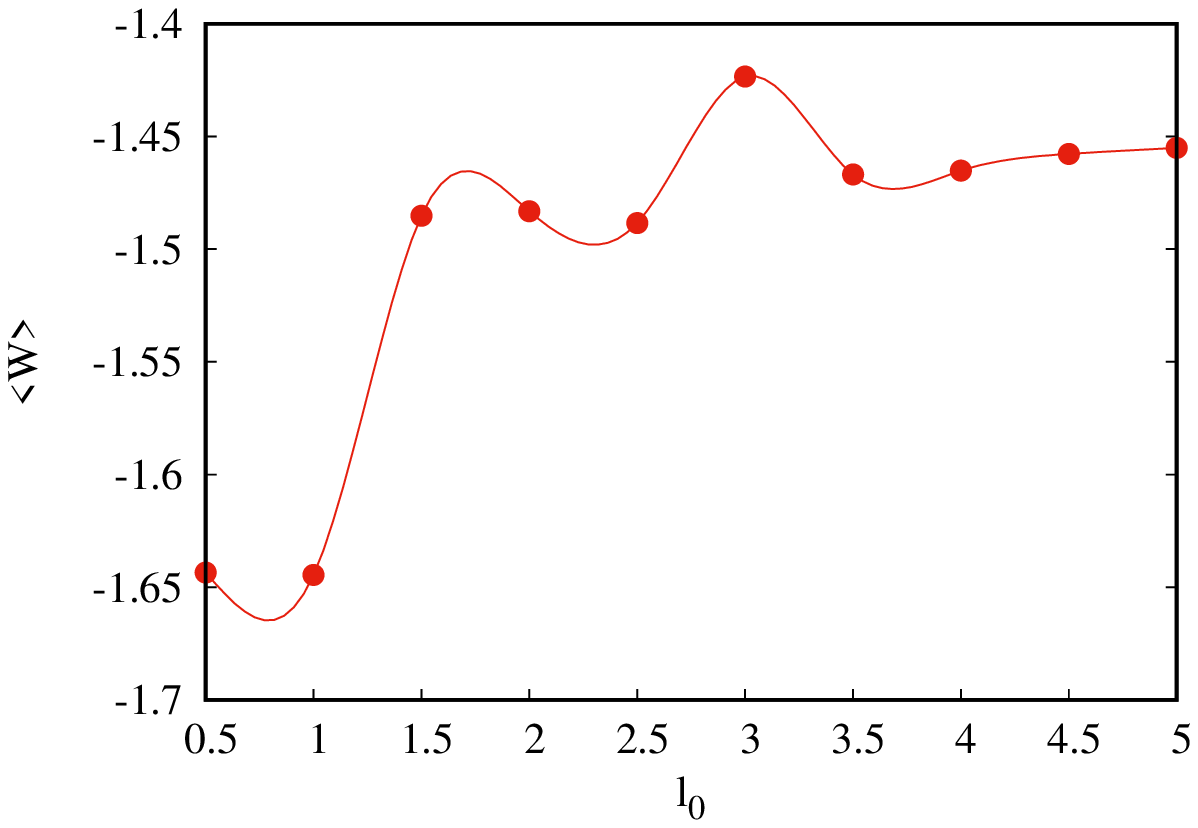}
        \caption{}
    \end{subfigure}
    \begin{subfigure}{0.45\textwidth}
        \includegraphics[width=\textwidth]{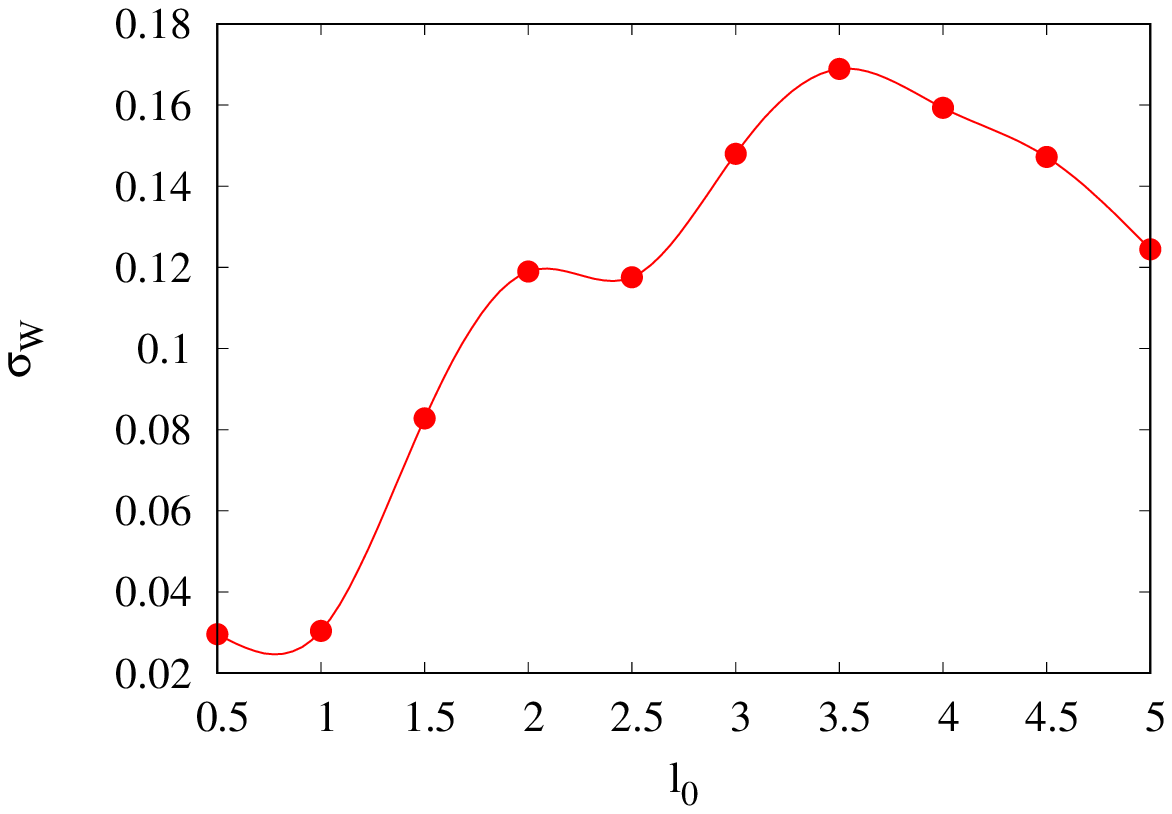}
        \caption{}
    \end{subfigure}
    \caption{(a) Mean work as a function of $l_0$. Parameters used: $N=50$, $f_{0}^e=20$ $f_{0}^c=10,~T_h=T_c=0.001$, $\alpha=1$. (b) Variance in work as a function of $l_0$ for the same set of parameters.}
    \label{fig:W_l0}
\end{figure}

Next, we discuss the dependence of the mean extracted work on the \emph{radius of neighbourhood} $l_0$ that is used by each particle to compute the mean direction of its neighbours (see discussion below Eq. \eqref{eq:ActiveForce}). This is another parameter characterizing the nature of activity - whether each particle is adjusting its motion by looking at particles in immediate vicinity or is looking at particles that are further off before deciding the direction along which it would proceed. Fig. \ref{fig:W_l0}(a) shows a sharp decline in the value of mean extracted work when $l_0$ increases from 1 to 1.5, beyond which it stays at a somewhat constant level with small fluctuations. The mean work eventually saturates when $l_0$ becomes of the order of the linear size of the cluster.
Just as in the case of variation with angular noise, the dependence of the variance of work on $l_0$ also shows non-monotonicity, as shown in Fig. \ref{fig:W_l0}(b). There is a peak at $l_0 \approx 3.5$, where the reliability of the engine is the least. The fluctuations in the work output becomes least close to $l_0\approx 0.5$, which shows that the output of the engine becomes more predictable when the radius of neighbourhood becomes smaller.

 Overall, the parameter that has the most significant effect on the extracted work turns out to be the difference between the activity strengths during the expansion and compression strokes. The choice of $l_0$ also alters the work output, but to a much smaller degree.
 As functions of $\alpha$ and $l_0$, we found that the work variances show non-monotonic behaviour, implying that the reliability of the engine is minimum within a small range of the concerned parameter.

\begin{figure}[h!]
    \centering
    \includegraphics[width=0.5\textwidth]{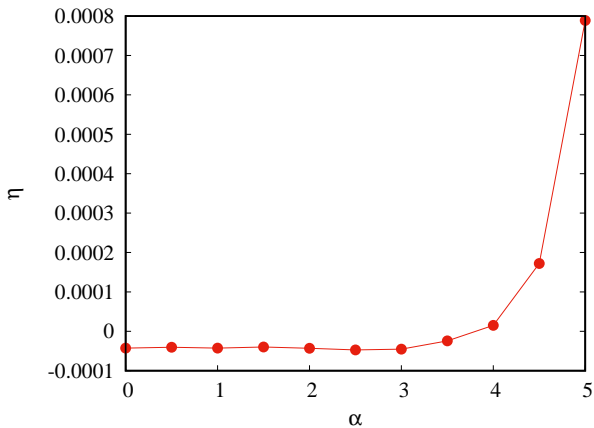}
    \caption{Plot showing the variation of efficiency ($\eta$) with angular noise ($\alpha$). The parameters are: $N=50$, $l_0=2$, $f_{0}^e=20$, $f_{0}^c=10$,~$T_h=T_c=0.001$.}
    \label{fig:eff_alpha}
\end{figure}

 \paragraph{Efficiency of the engine:}
In Fig. \ref{fig:eff_alpha} we now study the dependence of the efficiency of the engine on the angular noise. We find in our simulations that the magnitude of the heat exchanged with the bath during expansion, as computed using Eq. \eqref{eq:HeatDissipated} and \eqref{eq:InternalEnergyChange}, are several orders of magnitude larger than the extracted work. Furthermore, the sign of the heat exchanged is \emph{positive} in the entire range, so that the efficiency $\eta$ becomes negative in the range $\alpha<4$ where the work is being extracted. This implies that heat is being dissipated instead of being absorbed during the expansion stroke. This unusual result is not ruled out on physical grounds, since the entire engine is working in presence of a single-temperature bath, and the results expected for a normal two-bath engine may not be valid in this case. Furthermore, there are possibly a large number of degrees of freedom through which the system dissipates heat, thus leading to a large dissipation \cite{Barato2022}.  A negative value of efficiency is difficult to be interpreted physically, and Fig. \ref{fig:eff_alpha} has been provided only to bring out this anomaly. Any further significance of the efficiency as obtained from the aforementioned calculations will not be pursued in this article. Rather, we would like to state the study of efficiency as an open problem that needs more careful considerations.

\section{Conclusions}
 
In this work, we have studied an active thermal engine, whose working substance consists of  self-propelling and self-aligning particles constrained to move in two dimensions. They are confined in a spherically symmetric tan-hyperbolic potential. The steepness of the walls of the trapping potential as well as the activity of the particles are varied periodically with time.
 We find that due to the time-dependence of the potential, the proximity of the cluster to the walls also change periodically. Steeper the wall, closer is the cluster to it, and \textit{vice versa}.

We have defined the thermodynamic observables as per the prescription of stochastic thermodynamics in active systems \cite{sei19_prx}. The activity during the expansion step is kept higher than that in the compression step. Several interesting observations ensued. We found that the extracted work of the engine becomes higher with increased ratio of activities in the two halves of the engine cycle. The strength of angular noise also significantly alters the extracted work. The variances in work show non-monotonic behaviour as a function of the angular noise strength, with the peak being at $\alpha \approx \pi$. Such non-monotonic variations in the variance occurs even with change in the Vicsek radius. The peak can be interpreted as the region around which there are higher fluctuations around mean work, entailing higher errors in the mean observables like power and efficiency.

Due to possible hidden degrees of freedom, the internal energy change becomes dependent on the phase space trajectory. However, the definitions of stochastic thermodynamic observables of active systems are not free from ambiguities, and the different conventions used in literature are described in \cite{Fodor2022}. Due to the non-trivial nature of the active engine running in contact with a single heat bath, possibly arising from hidden degrees of freedom, the conventional definitions of efficiency shows deviations from the expected behaviour.

\section{Acknowledgments}
AS thanks the start-up grant from University Grants Commission (UGC) via UGC Faculty recharge program (UGCFRP) and the Core Research Grant (CRG/2019/001492) from SERB, India. AS and SL thank ICTS, India, where a majority of these discussions had taken place.


\begin{thebibliography}{10}

\bibitem{Carnot1824}
Sadi Carnot.
\newblock Reflections on the motive power of fire, and on machines fitted to
  develop that power.
\newblock {\em Paris: Bachelier}, 108:1824, 1824.

\bibitem{Carnot1897}
Sadi Carnot.
\newblock {\em Reflections on the Motive Power of Heat: From the Original
  French of N.-L.-S. Carnot}.
\newblock John Wiley, 1897.

\bibitem{Callen}
H.~B. Callen.
\newblock {\em Thermodynamics and an introduction to thermostatistics}.
\newblock John Wiley \& Sons, 2006.

\bibitem{Spudich1972}
James~A Spudich, Hugh~E Huxley, and John~T Finch.
\newblock Regulation of skeletal muscle contraction: \uppercase{II}. structural
  studies of the interaction of the tropomyosin-troponin complex with actin.
\newblock {\em J. Mol. Biol.}, 72(3):619--632, 1972.

\bibitem{Spudich1994}
James~A Spudich.
\newblock How molecular motors work.
\newblock {\em Nature}, 372(6506):515--518, 1994.

\bibitem{Alberts}
B.~Alberts, A.~Johnson, J.~Lewis, M.~Raff, K.~Roberts, P.~Walter, and
  N.~Chaffey.
\newblock {\em Molecular biology of the cell}.
\newblock Oxford University Press, 4th edition, 2003.

\bibitem{Astumian2001}
R~Dean Astumian.
\newblock Making molecules into motors.
\newblock {\em Sci. Am.}, 285(1):56--64, 2001.

\bibitem{Astumian2002}
R~Dean Astumian and Peter H{\"a}nggi.
\newblock Brownian motors.
\newblock {\em Phys. today}, 55(11):33--39, 2002.

\bibitem{Ait-Haddou2003}
Rachid Ait-Haddou and Walter Herzog.
\newblock Brownian ratchet models of molecular motors.
\newblock {\em Cell Biochem. Biophys.}, 38:191--213, 2003.

\bibitem{Feynman1986}
Richard~P Feynman.
\newblock {\em The Feynman Lectures on Physics Vol l}.
\newblock Narosa, 1986.

\bibitem{sei08_epl}
T.~Schmeidl and U.~Seifert.
\newblock Efficiency at maximum power: An analytically solvable model for
  stochastic heat engines.
\newblock {\em EPL}, 81:20003, 2008.

\bibitem{Rana2014}
Shubhashis Rana, PS~Pal, Arnab Saha, and AM~Jayannavar.
\newblock Single-particle stochastic heat engine.
\newblock {\em Phys. Rev. E}, 90(4):042146, 2014.

\bibitem{Abah2014}
J.~Ro\ss{}nagel, O.~Abah, F.~Schmidt-Kaler, K.~Singer, and E.~Lutz.
\newblock Nanoscale heat engine beyond the carnot limit.
\newblock {\em Phys. Rev. Lett.}, 112:030602, January 2014.

\bibitem{Verley2014}
Gatien Verley, Massimiliano Esposito, Tim Willaert, and Christian Van~den
  Broeck.
\newblock The unlikely carnot efficiency.
\newblock {\em Nat. Commun.}, 5(1):4721, 2014.

\bibitem{bechinger2012}
Valentin Blickle and Clemens Bechinger.
\newblock Realization of a micrometre-sized stochastic heat engine.
\newblock {\em Nat. Phys.}, 8:143, 2012.

\bibitem{Koski2014}
Jonne~V Koski, Ville~F Maisi, Jukka~P Pekola, and Dmitri~V Averin.
\newblock Experimental realization of a szilard engine with a single electron.
\newblock {\em Proc. Natl. Acad. Sci. U.S.A.}, 111(38):13786--13789, 2014.

\bibitem{Abah2016}
Johannes Ro{\ss}nagel, Samuel~T. Dawkins, Karl~N. Tolazzi, Obinna Abah, Eric
  Lutz, Ferdinand Schmidt-Kaler, and Kilian Singer.
\newblock A single-atom heat engine.
\newblock {\em Science}, 352(6283):325--329, 2016.

\bibitem{Goold2019}
D.~von Lindenfels, O.~Gr\"ab, C.~T. Schmiegelow, V.~Kaushal, J.~Schulz, Mark~T.
  Mitchison, John Goold, F.~Schmidt-Kaler, and U.~G. Poschinger.
\newblock Spin heat engine coupled to a harmonic-oscillator flywheel.
\newblock {\em Phys. Rev. Lett.}, 123:080602, August 2019.

\bibitem{roldan2016}
I.~A. Martinez, E.~Roldan, L.~Dinis, D.~Petrov, J.~M.~R. Parrondo, and R.~A.
  Rica.
\newblock Brownian carnot engine.
\newblock {\em Nat. Phys.}, 12:67, 2016.

\bibitem{Garcia2016}
Marc Serra-Garcia, Andr\'e Foehr, Miguel Moler\'on, Joseph Lydon, Christopher
  Chong, and Chiara Daraio.
\newblock Mechanical autonomous stochastic heat engine.
\newblock {\em Phys. Rev. Lett.}, 117:010602, June 2016.

\bibitem{sei19_prx}
Patrick Pietzonka, \'Etienne Fodor, Christoph Lohrmann, Michael~E. Cates, and
  Udo Seifert.
\newblock Autonomous engines driven by active matter: Energetics and design
  principles.
\newblock {\em Phys. Rev. X}, 9:041302, 2019.

\bibitem{Saadeh2014}
Yamaan Saadeh and Dinesh Vyas.
\newblock Nanorobotic applications in medicine: current proposals and designs.
\newblock {\em Am. J. Robot. Surg.}, 1(1):4--11, 2014.

\bibitem{ajay2016_nature}
Sudeesh Krishnamurthy, Subho Ghosh, Dipankar Chatterji, Rajesh Ganapathy, and
  A.~K. Sood.
\newblock A micrometre-sized heat engine operating between bacterial
  reservoirs.
\newblock {\em Nat. Phys.}, 12:1134, 2016.

\bibitem{Saha_2019}
Arnab Saha and Rahul Marathe.
\newblock Stochastic work extraction in a colloidal heat engine in the presence
  of colored noise.
\newblock {\em J. Stat. Mech. Theory Exp.}, 2019(9):094012, sep 2019.

\bibitem{lahiri2020}
Aradhana Kumari, P.~S. Pal, Arnab Saha, and Sourabh Lahiri.
\newblock Stochastic heat engine using an active particle.
\newblock {\em Phys. Rev. E}, 101:032109, 2020.

\bibitem{Kumari2021}
Aradhana Kumari and Sourabh Lahiri.
\newblock Microscopic thermal machines using run-and-tumble particles.
\newblock {\em Pramana}, 95:205, Nov 2021.

\bibitem{Rana2019}
Shubhashis Rana, Md~Samsuzzaman, and Arnab Saha.
\newblock Tuning the self-organization of confined active particles by the
  steepness of the trap.
\newblock {\em Soft Matter}, 15(43):8865--8878, 2019.

\bibitem{Vicsek1995}
Tam{\'a}s Vicsek, Andr{\'a}s Czir{\'o}k, Eshel Ben-Jacob, Inon Cohen, and Ofer
  Shochet.
\newblock Novel type of phase transition in a system of self-driven particles.
\newblock {\em Phys. Rev. Lett.}, 75(6):1226, 1995.

\bibitem{Weeks1971}
John~D Weeks, David Chandler, and Hans~C Andersen.
\newblock Role of repulsive forces in determining the equilibrium structure of
  simple liquids.
\newblock {\em J. Chem. Phys.}, 54(12):5237--5247, 1971.

\bibitem{sek98}
K.~Sekimoto.
\newblock Kinetic characterization of heat bath and the energetics of thermal
  ratchet models.
\newblock {\em Prog. theor. phys., Suppl.}, 130:17, 1998.

\bibitem{sekimoto}
Ken Sekimoto.
\newblock {\em Stochastic Energetics}.
\newblock Springer, 2010.

\bibitem{sei12_rpp}
U.~Seifert.
\newblock Stochastic thermodynamics, fluctuation theorems and molecular
  machines.
\newblock {\em Rep. Prog. Phys.}, 75:126001, 2012.

\bibitem{ris}
H.~Risken.
\newblock {\em The Fokker-Planck Equation: Methods of Solutions and
  Applications}.
\newblock Springer, 1996.

\bibitem{Barato2022}
Arya Datta, Patrick Pietzonka, and Andre~C. Barato.
\newblock Second law for active heat engines.
\newblock {\em Phys. Rev. X}, 12:031034, Sep 2022.

\bibitem{Fodor2022}
{\'E}tienne Fodor, Robert~L Jack, and Michael~E Cates.
\newblock Irreversibility and biased ensembles in active matter: Insights from
  stochastic thermodynamics.
\newblock {\em Annu. Rev. Condens. Matter Phys.}, 13:215--238, 2022.

\end{thebibliography}
\end{document}